\documentstyle[11pt,newpasp,twoside,epsf]{article}
\def\edcomment#1{\iffalse\marginpar{\raggedright\sl#1\/}\else\relax\fi}
\reversemarginpar

\def\sub#1{_{\mathrm{#1}}}

\begin{document}
\title{The effect of grain drift on the structure of (Post--) AGB winds}
\author{Yvonne Simis, Carsten Dominik and Vincent Icke}
\affil{Sterrewacht Leiden, Postbus 9513, 2300 RA Leiden, The Netherlands}
\begin{abstract}
We have developed an implementation for the momentum transfer force in
numerical two fluid hydrodynamics. This form of the frictional
coupling between gas and grains is consistent with the microscopic
interactions between the two components. The coupling force gives rise
to a drift velocity of the grains with respect to the gas. We apply
this mechanism to the outflow of (Post-) AGB objects. Our numerical
hydrodynamics code calculates self consistently the dynamics of these
outflows, as well as the nucleation and growth of grains and
equilibrium chemistry of the gas. Grain nucleation and growth are
processes that depend strongly on the rate of gas--grain collisions.
Hence, the drift velocity becomes an important variable.
The tight connection between grain chemistry and drift
causes the system to become extremely sensitive to small changes in	
almost any parameter. This may be a cause for deviation from (spherical)
symmetry and structure.
\end{abstract}
\section{Dust driven winds}
Dust driven winds are powered by a fascinating interplay of radiation,
chemical reactions, stellar pulsations and atmospheric dynamics. As
soon as an AGB star's atmosphere develops sites suitable for the
formation of solid ``dust'' (i.e. sites with a relatively high density
and a low temperature) its dynamics will be dominated by the power of
the radiative force. Dust grains absorb stellar radiation efficiently
and experience a large radiation pressure. The momentum thus
acquired is partially transferred to the ambient gas by
frequent collisions.  The gas is then blown outward in a dense, slow
wind that can reach high mass loss rates.
\section{Numerical hydrocode}
\begin{figure}
\plotfiddle{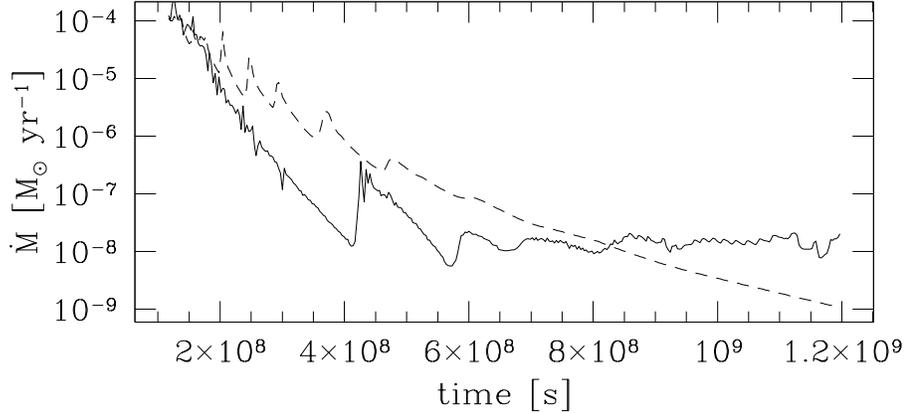}{4.5cm}{0}{60}{60}{-190}{-100}
\caption{Mass loss rates for thermal (dashed line) and drift driven grain chemistry.}
\end{figure}
We have written a numerical hydrodynamics code that self consistently
calculates a dust driven wind. In our code both gas and dust are
described by their own set of hydro equations (continuity,
momentum). Exchange of matter (nucleation and growth of grains) and
momentum (collisions) are taken into account in the source terms. The
time dependent continuity and momentum equations are numerically
solved using a two--step FCT/LCD algorithm (Boris 1976; Icke
1991). The abundances of H, H$_2$, C, C$_2$, C$_2$H, C$_2$H$_2$ and CO
in the gas are calculated using a simple equilibrium chemistry
(Dominik et al. 1990). Nucleation and growth of dust grains is
described by the moment method (Gail, Keller, \& Sedlmayr 1984; Gail
\& Sedlmayr 1988; Dorfi \& H\"ofner 1991).
\section{Two fluid hydrodynamics}
Two fluid hydrodynamics requires a careful implementation of the
momentum transfer term in terms of the drift velocity of the grains
with respect to the gas. It turns out that the expression for this
drag force that has been used before (e.g. Dominik 1992; Berruyer 1991; 
Kr\"uger, Gauger, \& Sedlmayr 1994) doesn't always
apply, in particular when or where grains have just started to form. Moreover
this expression just takes into account what we will call the
``macroscopic'' component of the drift velocity and doesn't incorporate
the contribution to the momentum transfer due to the radiative
acceleration between two subsequent collisions of a grain
(``microscopic drift''). We have derived a new implementation for the
momentum transfer from dust to gas, see Simis, Icke, \& Dominik (2000):
\begin{equation}
f\sub{drag} = n\sub d g\sub{rad} \frac{m\sub g}{m\sub g + m\sub d} 
\left(1 + \frac{v}{\sqrt{v^2 + 2\lambda g\sub{rad}} - v} \right)
\label{eq:fdrag}
\end{equation}
Here, $n\sub d$ is the number density of grains, $m\sub{d,g}$ is the mass
of a dust/gas particle, $\rho\sub{d,g}$ are mass densities, $g\sub{rad}$ is
the radiative acceleration of a grain, $v = v\sub d - v\sub g$ is the drift velocity,
$\lambda$ is the mean free path of a grain and 
$\Omega = \rho\sub g m\sub d - \rho\sub d m\sub g/\rho\sub g (m\sub g + m\sub d)$.
This expression takes into account
the contribution of the radiative acceleration between individual
gas--grain encounters as well. Especially when the mean free path of
the grains is large, this radiation pressure contribution is
important and may be the dominant factor in the momentum transfer.
The equilibrium drift velocity corresponding to Eq. (1) is (Simis et al. 2000)
\begin{equation}
v\sub{eq} = \sqrt{\frac{\Omega^2}{1-\Omega^2}2\lambda g\sub{rad}}
\label{eq:veq}
\end{equation}
This formalism will in the future allow to treat the coupling force
also at non--equilibrium drift speeds. For the current calculations we
have still assumed that the grains reach equilibrium speed quickly
enough for the rate of momentum transfer to be given by $v = v\sub{eq}$, i.e.
\begin{equation}
f\sub{drag} = n\sub d g\sub{rad} \frac{\rho\sub g}{\rho\sub g + \rho\sub d} 
\end{equation}
\section{The effect of drift on the dynamics of the wind}
\begin{figure}
\plottwo{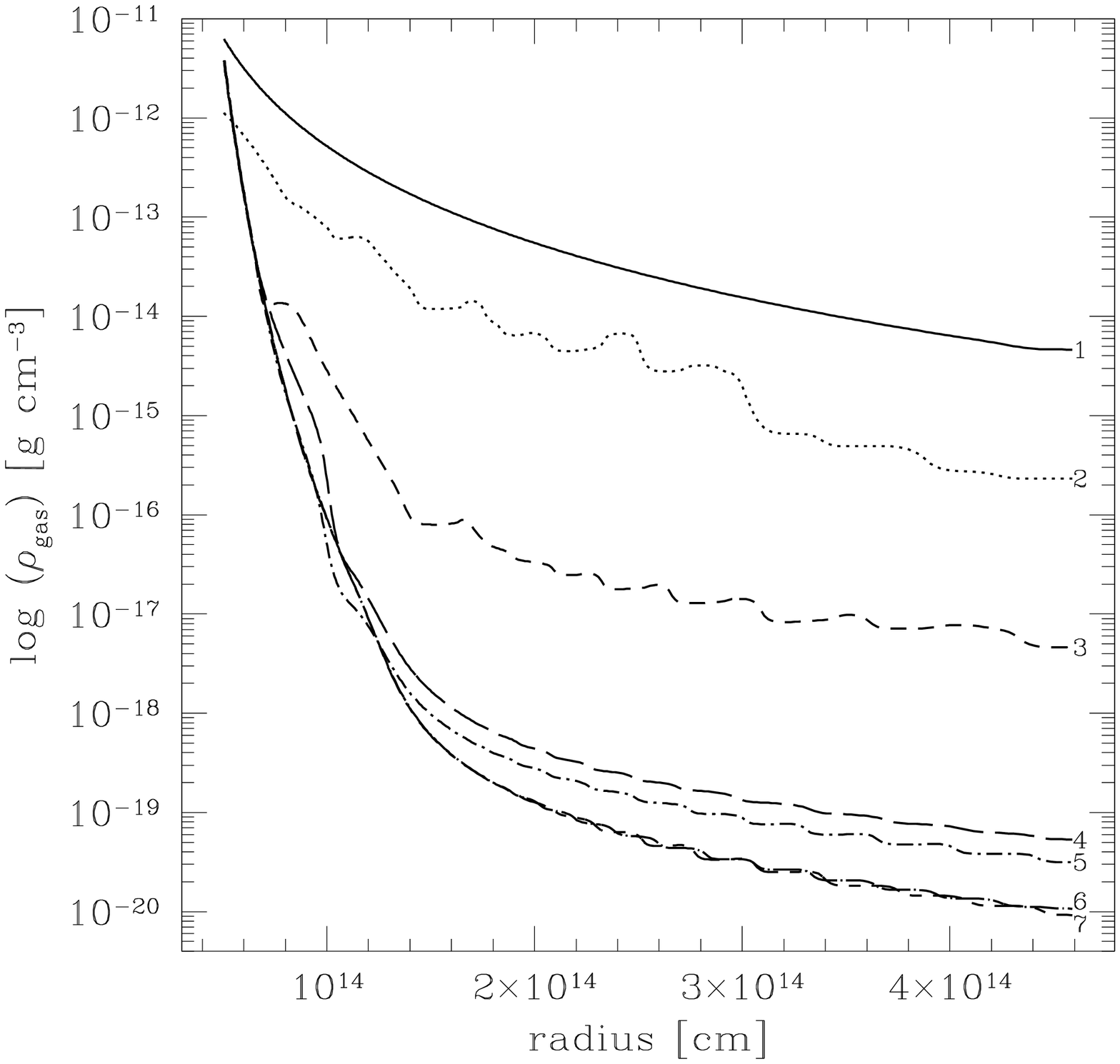}{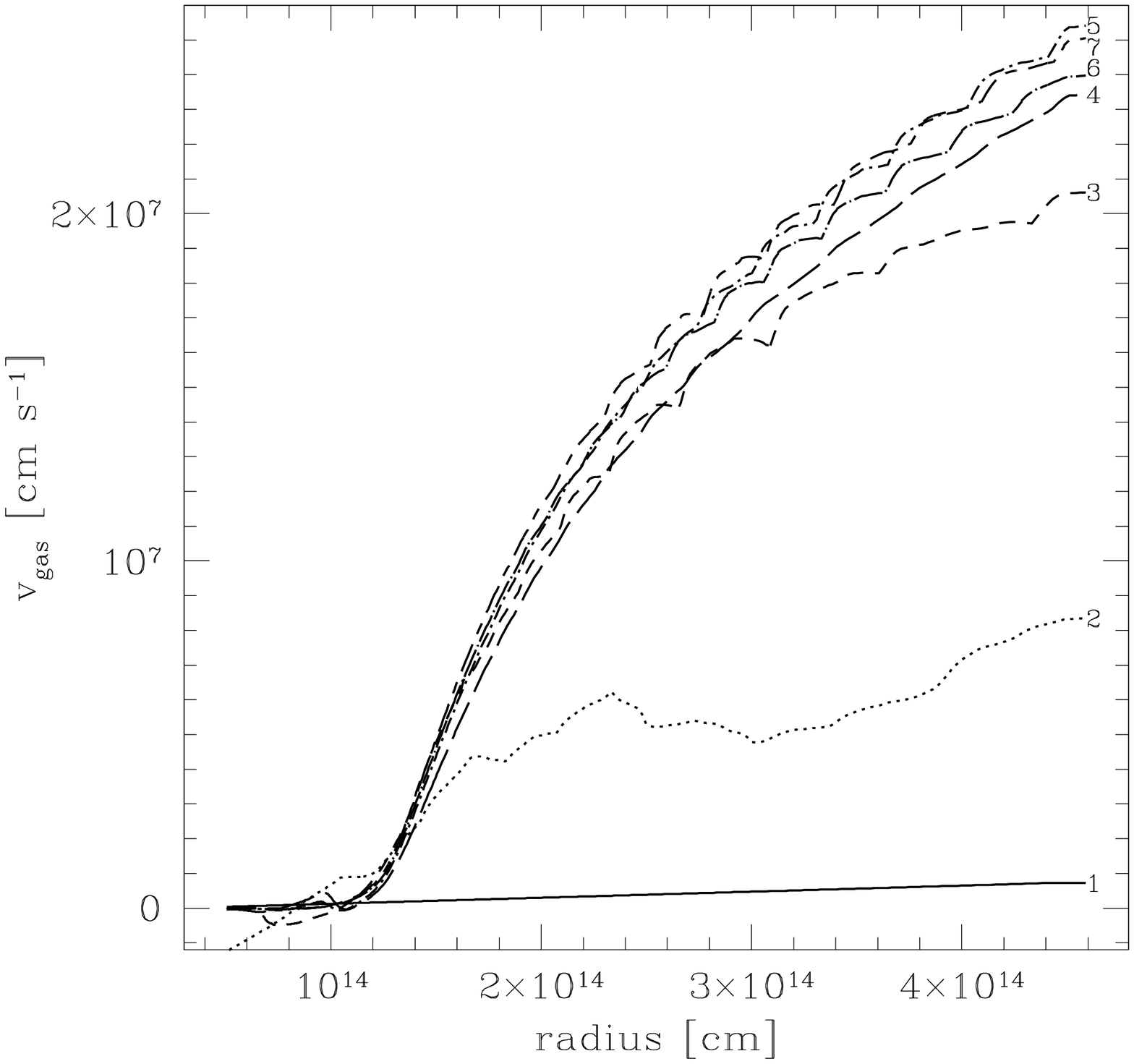}\\
\plottwo{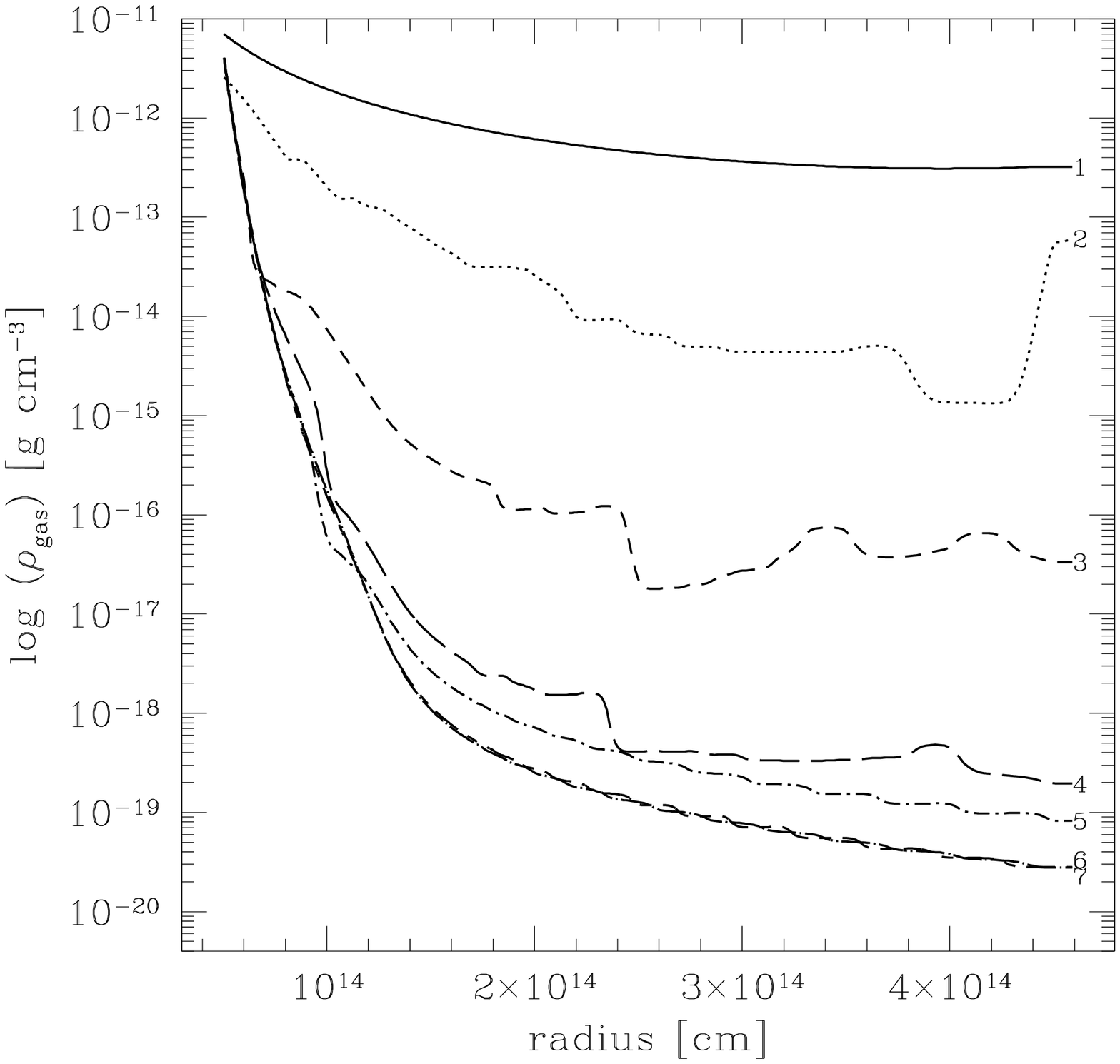}{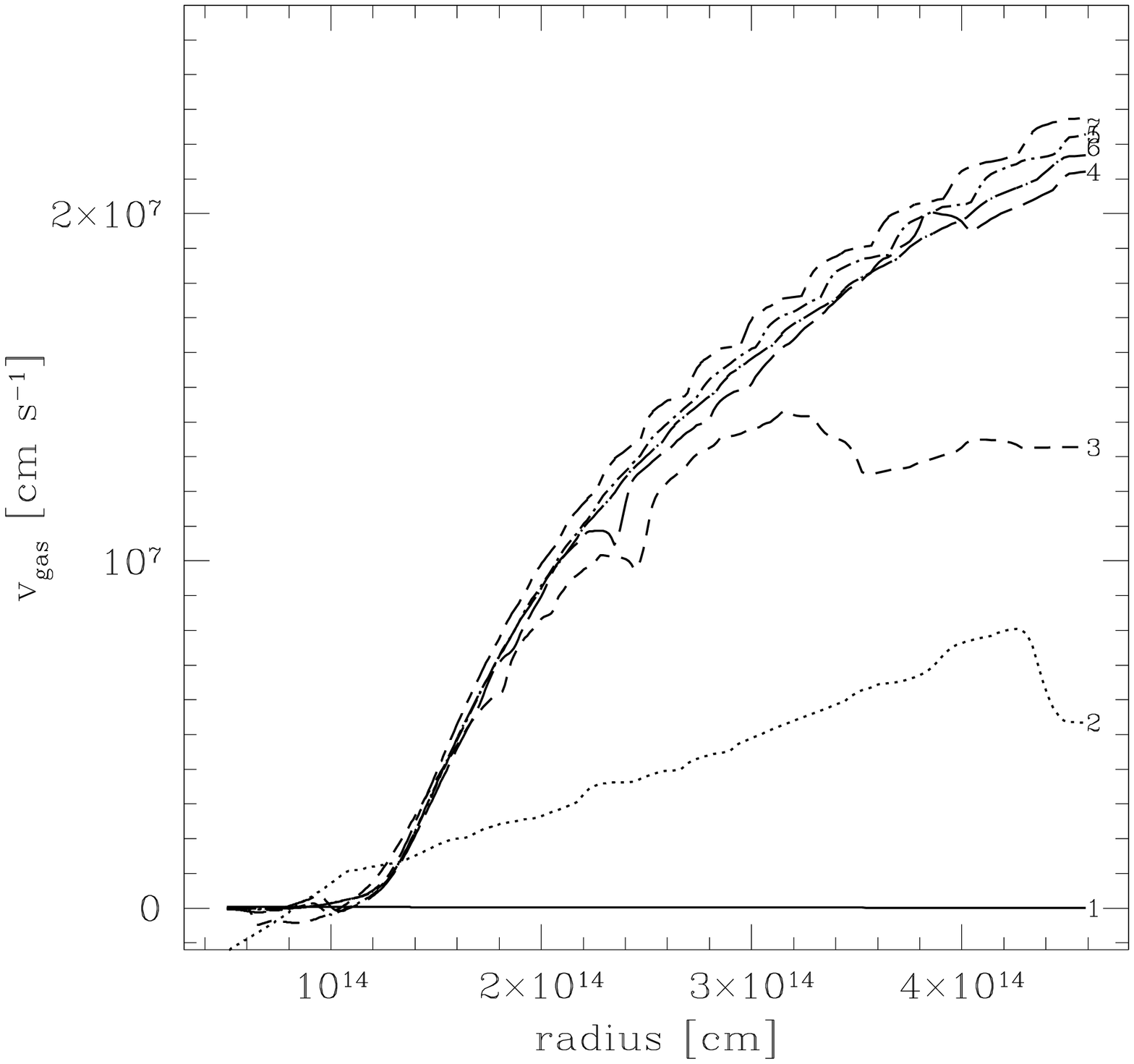}
\caption{Density and velocity structure of simulated polar (upper two plots) and
equatorial (lower two plots) outflow. Time after start of the wind is $0.0 s$ (1), $0.57 \times 10^8 s$ (2), $2.07 \times 10^8 s$ (3), $3.57 \times 10^8 s$ (4), $5.07 \times 10^8 s$ (5), $6.57 \times 10^8 s$ (6), $8.07 \times 10^8 s$ (7)}\mbox{}
\label{fig:rotating}\\
\end{figure}
The rates of grain nucleation and growth depend on the velocity with which
grains and gas particles collide. In absence of drift, gas--grain
collisions are thermally driven. Now that we have an expression for
the drift velocity that is consistent with the micro--dynamics we can
take into account the effect of drift on grain nucleation and growth.
Since the drift velocity, through the radiation pressure, depends on
the number density and size spectrum of the grains we are now dealing
with a very strong coupling between the dynamics and the chemistry
rates in the outflow. When drift is taken into account, and in
absence of sputtering, grains become
bigger and more abundant than in the case when only thermal collisions
are considered. In the case of drift driven grain chemistry, a wind with a
terminal velocity and a mass loss rate that fluctuate around a
constant value establishes itself. In the case of purely thermally driven
grain chemistry, the gas outflow velocity and mass loss rate keep
decreasing. At the same time dust continues to flow out at a high
rate. This is illustrated in Figure 1, in which the mass loss history of
an object with $M_* = 1 M_\odot$, $T_* = 2200\ K$, $L_* = 1.0 \times
10^4 L_{\odot}$ and $\epsilon_C/\epsilon_O = 2$ is shown.\\
Due to the strong coupling of chemistry and dynamics, a small change
in the parameters or flow variables may result in large changes in the
flow. To illustrate this, we compare the outflows of a non--rotating
object and an object with a rotation period of 50 years. The
non--rotating object may be interpreted to represent the outflow in
the polar direction and the rotating object to represent the outflow
in the equatorial plane of an AGB star with a 50 year rotation
period. The rotation is simulated by suitable adjustment of the
effective gravity. Figure 2 shows the dynamical evolution of the
density and velocity structure of the outflows. It turns out that the
mass loss rate in the equatorial plane is twice the mass loss rate in
the polar direction. The velocity in the polar direction is higher and
the density is lower than in the equatorial
plane.
One may conclude that this zeroth order model of a rotating AGB
object indicates that the initial polar to equatorial density gradient
gives rise to a significant difference in mass loss rate, which may
lead to a disk like structure. More generally this illustrates the
tight coupling between dynamics and chemistry.
%
%
%
%
%
%
%
%

\end{document}